\documentclass[%
reprint,
amsmath,amssymb,
pra,superscriptaddress
]{revtex4-1}

\usepackage[utf8]{inputenc}
\usepackage{graphicx}
\usepackage{dcolumn}
\usepackage{bm}
\usepackage{mathrsfs}
\usepackage{amsfonts}
\usepackage{physics}
\DeclareMathOperator*{\argmax}{argmax}

\begin{document}
	
	\title{Approaching Quantum Limited Super-Resolution Imaging without Prior Knowledge of the Object Location}
	
	\author{Michael R Grace}
	\affiliation{College of Optical Sciences, University of Arizona, Tucson, AZ 85719, USA}
	\author{Zachary Dutton}
	\affiliation{Physical Sciences and Systems, Raytheon BBN Technologies, Cambridge, MA 02138, USA}
	\author{Amit Ashok}
	\affiliation{College of Optical Sciences, University of Arizona, Tucson, AZ 85719, USA}
	\author{Saikat Guha}
	\affiliation{College of Optical Sciences, University of Arizona, Tucson, AZ 85719, USA}
	
	\begin{abstract}
		A recently identified class of receivers which demultiplex an optical field into a set of orthogonal spatial modes prior to detection can surpass canonical diffraction limits on spatial resolution for simple incoherent imaging tasks. However, these mode-sorting receivers tend to exhibit high sensitivity to contextual nuisance parameters (e.g., the centroid of a clustered or extended object), raising questions on their viability in realistic imaging scenarios where little or no prior information about the scene is available. We propose a multi-stage passive imaging strategy which segments the total recording time between different physical measurements to build up the required prior information for near quantum-optimal imaging performance at sub-Rayleigh length scales. We show via Monte Carlo simulations that an adaptive two-stage scheme which dynamically allocates the total recording time between a traditional direct detection measurement and a binary mode-sorting receiver outperforms idealized direct detection alone for simple estimation tasks when no prior knowledge of the object centroid is available, achieving one to two orders of magnitude improvement in mean squared error. Our scheme can be generalized for more sophisticated imaging tasks with multiple parameters and minimal prior information.
	\end{abstract}
	\maketitle
	
	\section{Introduction}
	\label{sec:introduction}
	
	The chief goal of quantitative imaging is to infer one or more properties of an object from an optical field emanating from it. In the context of passive optical imaging, this inference is typically made via a measurement process known as direct detection: the intensity of the optical field collected by an imaging system is measured at an image plane, generating an electrically encoded spatial mapping of the object which is subsequently digitally processed and analyzed. Due to diffraction from a finite aperture, resolution degradation for direct detection occurs for object features at high spatial frequencies. This small length scale regime relative to the system point spread function is commonly termed the ``sub-Rayleigh" regime, in reference to the well known Rayleigh criterion for distinguishability of point sources when using a diffraction limited imaging system and a direct detection measurement \cite{Rayleigh1879}. 
	
	Recently, the degree to which the choice of optical measurement affects quantitative imaging performance has received considerable attention. Diffraction impedes the precision of parameter estimation through the introduction of photon shot noise, which is added during the physical detection process by which the optical field collected by the aperture is converted to an electrical signal \cite{Gagliardi1976}. While shot noise cannot be entirely avoided, pre-detection optical transformations can be used to dispose the collected light in a more favorable manner to the inevitable shot noise introduced when the field is detected. As a result, it is sometimes possible to find an optical measurement, the description of which encompasses both the physical detection process and any pre-detection optical transformations applied to the field, that outperforms direct detection for certain imaging tasks. The optimal measurement for extracting spatial information from a weak, incoherent field always involves detection of the field intensity, since such fields exhibit no exploitable phase correlations \cite{Helstrom1976}, but the variety of measurements allowed by quantum mechanics is almost unlimited even before considering the plethora of available digital post-processing techniques. Optical measurement schemes for passive imaging employ techniques such as PSF engineering \cite{HELL1994,Kao1994,Juette2008,Pavani2009,Oti2005,Paur2018,Paur2019}, pre-detection interferometry \cite{Hell1994b,Gustafsson1999,Backlund2018a,Monnier2003,Nair2016b}, and multi-photon coincidence measurements \cite{Parniak2018}.
		
	Tsang et al. \cite{Tsang2016a} considered a class of measurements known as spatial-mode demultiplexing (SPADE), deriving inspiration from a quantum information theoretic analysis of the problem of estimating the separation between two incoherent point sources. SPADE refers to any passive measurement which projects the incident optical field onto some orthonormal basis of spatial modes in the optical domain. This projection is achieved by an optical mode sorting device, which couples each spatial mode of the image plane optical field to a designated intensity resolving detector. Such mode sorters have been experimentally demonstrated using interferometry \cite{Abouraddy2011,Zhou2017,Gu2018a,Malhotra2018}, holography \cite{Berkhout2010a, Mirhosseini2013a, Malik2014a, Trichili2016a}, multi-plane phase transformations \cite{Morizur2010a, Fontaine2019} and mode-selective phase shifts accessed by polarization elements \cite{Zhou2018b,Fu2018}. For simple, sub-Rayleigh parameter estimation \cite{Tsang2016a, Paur2016a, Tham2017, Hassett2018, Dutton2019, Tsang2019} and hypothesis testing \cite{Lu2018} tasks, simplified versions of SPADE measurements that use only two intensity resolving detectors surpass direct detection and approach the ultimate limits imposed by quantum mechanics \cite{Helstrom1976}.
	
	Unfortunately, mode sorting measurements can suffer from increased sensitivity to variations in environmental and/or experimental conditions compared with image-plane direct detection. One important factor that has been noted as a potential source of additional error for SPADE or other quantum inspired measurements is misalignment between the spatial centroid of the object being imaged and the origin of the mode basis associated with the optical sorting device \cite{Parniak2018, Tsang2016a, Chrostowski2017, Rehacek2018, Zhou2019, Prasad2019}. Optical misalignment can be a significant problem when the location of the object is initially unknown or when the imaging system exhibits appreciable alignment error. This issue is relevant in applications of interest such as fluorescence microscopy, astronomy, remote sensing, and tactical imaging. In these contexts, the centroid of a clustered or extended object can often be thought of as a \emph{nuisance parameter}, an unneeded piece of information for the specified imaging task that nonetheless affects the performance of the measurement. 
	
	In this work we propose and analyze an adaptive two-stage detection scheme which divides the total integration time between two physical measurements: a preliminary direct detection measurement to estimate the object centroid followed by a spatial mode-sorting measurement aligned to the centroid estimate. We analyzed the performance of our proposed receiver for two sub-Rayleigh estimation tasks, that of estimating the separation between two point sources and estimating the length of an extended object. Both of these tasks have been previously studied for the case in which the object centroid position is exactly known \textit{a priori} \cite{Tsang2016a,Dutton2019}, and some preliminary work has considered the unknown centroid case, discussing sequential measurement strategies in the limiting cases of long integration time \cite{Chrostowski2017} or small length scale \cite{Rehacek2018,Zhou2019,Prasad2019}. In contrast to previous studies, we consider finite integration time and no prior information on the centroid or scale parameters, particularly applicable conditions for biological and terrestrial imaging scenarios. We find via Monte Carlo simulation that our two-stage receiver outperforms a receiver that allocates the entire integration time to ideal direct detection, even in the absence of any prior knowledge of the object parameters. In many regimes, particularly with higher photon flux, the performance of the two-stage receiver approaches that of an initially aligned mode sorter, representing the optimal estimation precision allowed by quantum mechanics. We find that this improvement is realized with a straightforward strategy of simply dividing the integration time into two equal halves. However, we further show that adaptively tuning the allocation of integration time between the two measurement stages leads to additional receiver improvement, fully exploiting the merits of the mode-sorter. 

	\section{Background and Notation}
	\label{sec:theory}
	
	\subsection{Passive Measurements for Incoherent Imaging} \label{sec:measurements}
	
	Consider a passive, lossless imaging system collecting light from a far-field, quasi-monochromatic, incoherently radiating object. For a static, 1D object, the emitted optical field can be characterized by the normalized radiant exitance $m(x\vert\Theta)\geq0$ of the object, where $\int_{-\infty}^{\infty}m(x\vert\Theta)dx=1$ over the transverse object plane coordinate $x~\in~\mathbb{R}$ and where $\Theta$ is a vector of spatial parameters conditioning the radiant exitance profile. For simplicity, assume spatially invariant linear optics which map the object plane with unit magnification onto a conjugate image plane. Following standard far-field diffraction theory \cite{Goodman2005}, the imaging system is characterized by its coherent point spread function (PSF) $\psi(x)$, the 1D spatial Fourier transform of the effective aperture function. For weakly radiating incoherent sources from which less than one photon is detected per coherence time interval, an intensity resolving measurement of the optical field (e.g., excluding multi-photon interference measurements \cite{Parniak2018}) can be modeled as generating i.i.d. single photon detection events \cite{Tsang2016a} with Poisson temporal statistics \cite{Mandel1959, Basano2005}.
	
	The most widespread measurement for incoherent optical imaging is image-plane direct detection using a focal plane array (FPA) of intensity resolving detectors. An FPA can consist of a physical detector array, such as a CCD or CMOS camera, or a virtual array, such as that generated using a raster scan with a single photodetector (e.g., confocal microscopy). For an ``ideal" shot noise limited direct detection receiver (exhibiting unity quantum efficiency, no dark current or read noise, infinite spatial extent in the image plane and infinite spatio-temporal bandwidth), individual 1D measurement outcomes correspond to image-plane arrival positions $x_i\in \mathbb{R}$ of single photons. As a result, the single-photon outcome probability density is given by the image-plane intensity distribution 
	\begin{equation}
	\Psi(x_i\vert\Theta)=\int_{-\infty}^{\infty}m(x\vert\Theta)\abs{\psi(x_i-x)}^2dx,
	\label{eq:image_plane}
	\end{equation}
	the convolution between the object radiant exitance and the system PSF \cite{Goodman2005}. Ideal direct detection is a shift invariant measurement, meaning that the alignment of the detector array within the image plane has no effect on the form of $\Psi(x_i\vert\Theta)$. Over $n$ i.i.d.~single photon arrival events, the joint probability density for the set of direct detection arrival positions $\{x_i\}_{i=1}^n$ becomes
	\begin{equation}
	P_{\textrm{D}}\big(\{x_i\}_{i=1}^n\vert\Theta\big)=\prod\limits_{i=1}^n \Psi(x_i\vert\Theta).
	\label{eq:direct_joint_probability}
	\end{equation}
	
	A spatial mode demultiplexing (SPADE) measurement detects the intensity of the optical field after it is projected onto a discrete, orthonormal basis of spatial modes in the optical domain. Let the amplitude profile of the $q^{\textrm{th}}$  such spatial mode be defined by the real function $f_q(x)$, so that $\int_{-\infty}^{\infty}f_q(x)f_{q'}(x)dx=\delta_{qq'}$, where $\delta_{qq'}$ is the Kronecker delta function. Unlike direct detection, the measurement outcome distribution of even an ideal SPADE receiver may depend on optical alignment, since the optical projection is performed on a spatial mode basis whose transverse alignment is registered to a physical mode sorting device~\cite{Tsang2016a}. If $\phi_{\textrm{0}}$ denotes the transverse position where the optical axis of the mode sorter intersects the image plane, then the probability of photon detection in the $q^{\textrm{th}}$ spatial mode, conditioned on the arrival of a single photon, is given by \cite{Tsang2018a}
	\begin{equation}
	g(q\vert\Theta)=\int\limits_{-\infty}^{\infty}m(x\vert\Theta)\abs{\int\limits_{-\infty}^{\infty}\psi(s-x)f_q(s-\phi_{\textrm{0}})ds}^2 dx,
	\label{eq:BSPADE_model}
	\end{equation} 
	where $s$ is a real-valued spatial variable in the image plane. 
	
	A binary SPADE (BSPADE) device, a special case of SPADE, couples a single ``target" spatial mode to one detector and aggregates the complement of this mode into a second detector, resulting in two possible single-photon detection outcomes \cite{Tsang2016a}. For a set of $n$ such outcomes, the two integers $n$ and $k$ form a sufficient statistic, where $k$ is the number of photons recorded by the target detector. For i.i.d.~$q$-BSPADE measurements, where the $q^{\textrm{th}}$ spatial mode is designated as the target mode, the joint probability distribution governing $n$ recorded photons is given by the binomial distribution 
	\begin{equation}
	P_{\textrm{B}}(k,n\vert\Theta)=g(q\vert\Theta)^{k}\big(1-g(q\vert\Theta)\big)^{n-k}.
	\label{eq:BSPADE_joint}
	\end{equation}
	
	\subsection{Optical Misalignment as a Nuisance Parameter} \label{3_Misalignments}
	For many quantitative imaging tasks, transverse optical misalignment plays the role of a nuisance parameter which may degrade receiver performance while providing no useful scientific information on its own \cite{VanTrees2013}. Suppose that a transverse register $\phi$ on the object is known to correspond to the optimal (B)SPADE mode sorter alignment for some designated imaging task. We then define the nuisance parameter $\xi=\phi-\phi_{\textrm{0}}$ as the real-valued spatial offset of the optimal alignment with respect to the true mode sorter optical axis $\phi_{\textrm{0}}$. 
	
	Here, we model the misalignment $\xi$ using the sum  $\xi=\xi_{\textrm{P}}+\xi_{\textrm{S}}$. In order to spatially align an optical receiver, a point at the object plane must be chosen as the intended alignment position, which we denote as $\hat{\phi}$. The first contribution to optical misalignment, $\xi_{\textrm{P}}=\phi-\hat{\phi}$, arises from imperfect prior knowledge of the object location and equals the displacement of the optimal object-plane alignment position from the intended mode sorter alignment for the measurement. The second term, $\xi_{\textrm{S}}=\hat{\phi}-\phi_{\textrm{0}}$, reflects any additional offset between the intended alignment and the physically implemented alignment due to systematic pointing error of the imaging system. 
	
	In a well calibrated imaging context, the static misalignment $\xi$ is likely to be best described as a random variable with a known prior probability distribution $p(\xi)$. If the characterized physical misalignment processes $\xi_{\textrm{P}}$ and $\xi_{\textrm{S}}$ are statistically independent random variables, then $p(\xi_{\textrm{A}})=p(\xi_{\textrm{P}})\ast p(\xi_{\textrm{S}})$ is given by a convolution of the component prior distributions. In the present analysis, we assume for simplicity that $\xi_{\textrm{S}}$ is the result of unbiased, symmetric misalignment processes inherent to the imaging receiver and model it as a Gaussian distributed random parameter with zero mean and real-valued variance $\sigma_{\textrm{S}}^2$. For a given imaging trial, a distribution $p(\xi_{\textrm{P}})$ can be constructed using any available \textit{a priori} information on the optimal object-plane alignment position $\phi$ and/or the results of any preliminary measurements thereof. The magnitude of each of these sources of optical misalignment will vary depending on the imaging context. For example, if the location of the object is not known with adequate precision before the measurement, the distribution $p(\xi_{\textrm{P}})$ is likely to be broad, while poor optomechanical precision could result in large $\sigma_{\textrm{S}}$. 
	
	\section{Results}
	\label{sec:results}
	
	\subsection{Point Source Separation Estimation Task}
	\label{sec:tasks}
	
\begin{figure}[htb]
		\centering
		\includegraphics[width=\linewidth]{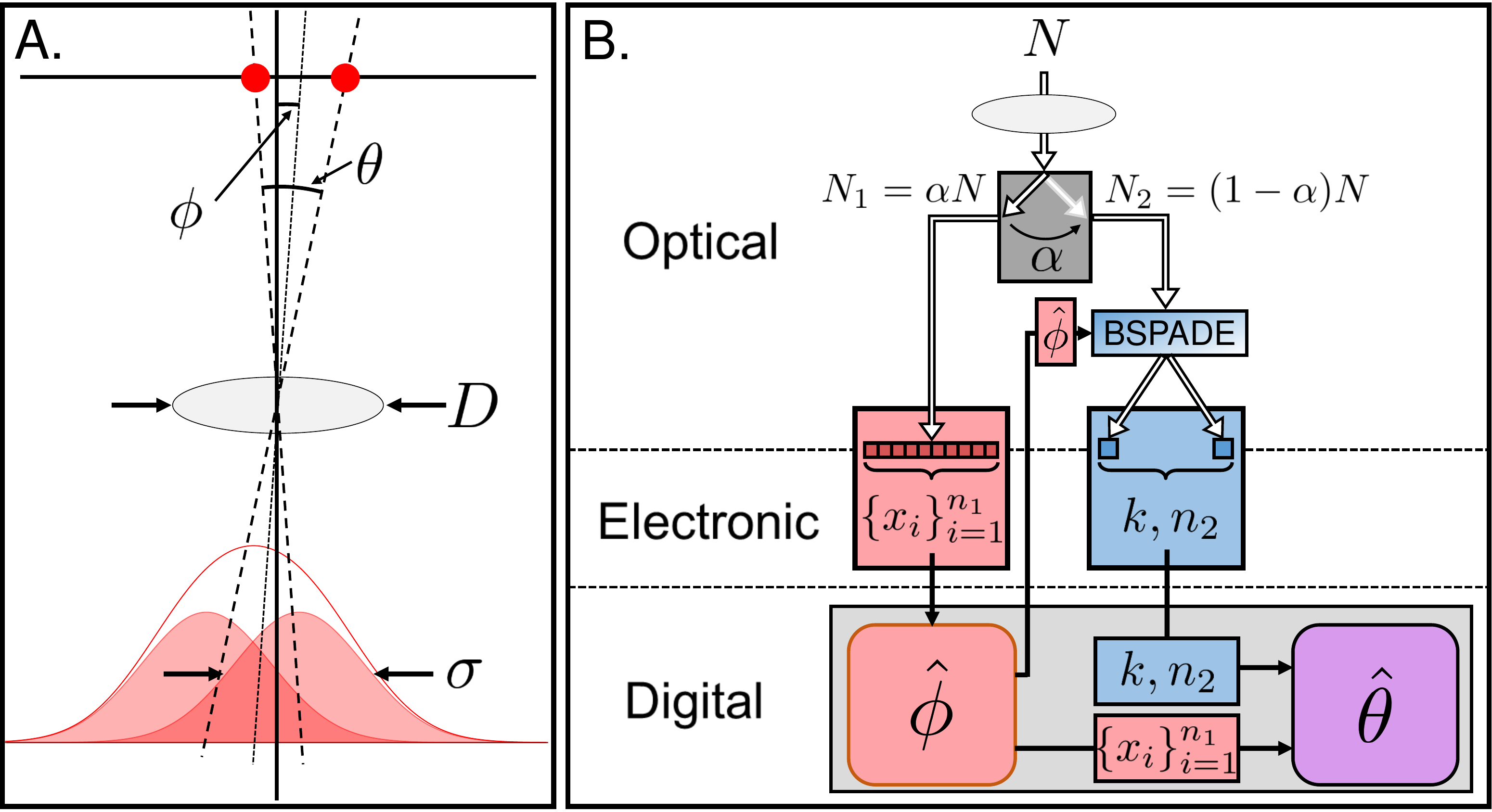}
		\caption{A. Image plane intensity distribution from a two point source object with angular centroid $\phi$ and angular separation $\theta$ imaged with a Gaussian aperture. B. Process flow for the proposed two-stage receiver design. The total mean photon number $N$ is divided between a direct detection measurement and an aligned BSPADE measurement according to the free design parameter $\alpha$. A direct detection measurement is used to estimate the object centroid and physically align a BSPADE measurement to the centroid estimate $\hat{\phi}$ in a feed-forward routine.
		}
		\label{fig:2Stage}
	\end{figure}

	For the primary results of our work, we consider two identical, incoherent point sources with centroid $\phi\in\mathbb{R}$ and separation $\theta\in\mathbb{R}$ and seek to estimate $\theta$ with no prior knowledge of either parameter (Fig. \ref{fig:2Stage}A). Let the total available integration time for an imaging trial be $T$, during which the mean number of photons detected by the receiver is known to be $N$.  We model the spatial properties of the two point source object with the normalized radiant exitance
	\begin{equation}
	m(x\vert \theta, \phi)=\frac{1}{2}\big[\delta(x-\phi-\theta/2)+\delta(x-\phi+\theta/2)\big],
	\label{eq:PointSources_radiance}
	\end{equation}
	where $\delta(\cdot)$ is the Dirac delta function and $\theta>0$. With this task formulation, $\theta$ is the parameter of interest to be estimated while $\phi$ is a nonrandom but unknown nuisance parameter. 
	
	\subsection{Two Stage Receiver Design}
	\label{sec:Two-Stage_design}
	
	Our proposed two stage receiver, summarized in Fig. \ref{fig:2Stage}B, divides the total integration time $T$ of an imaging trial into two sequential measurement stages followed by digital post-processing.  Since the centroid $\phi$ of the two point sources is known to be the optimal (B)SPADE alignment position for estimation of the source separation \cite{Tsang2016a}, the receiver spends part of the integration time on a preliminary direct detection measurement to acquire prior information on the centroid location. A centroid estimate $\hat{\phi}$ will then be used as the intended alignment position for a second stage BSPADE measurement. The allocation of integration time between the two measurement stages is governed by the free design parameter $\alpha\in[0,1]$ so that the two measurements are active for times $\alpha T$ and $(1-\alpha) T$, respectively.
	
	In the first stage, a direct detection measurement records a Poisson distributed number of photons with mean $N_1=\alpha N$. For a particular imaging trial, let $n_1$ represent the number of photons recorded by the first stage at image-plane arrival positions $\{x_i\}_{i=1}^{n_1}$ distributed by Eq. \ref{eq:image_plane}. Utilizing the ``center of mass" estimator $\hat{\phi}=\frac{1}{n_1}\sum_{i=1}^{n_1} x_i$ as an approximation to the maximum likelihood (ML) estimator of the centroid (see Appendix \ref{apx:estimators} for additional derivations and details on estimators), the random variable $\xi_{\textrm{P}} = \phi-\hat{\phi}$ then corresponds to the error in the centroid estimate. Note that since $\hat{\phi}$ is unbiased, $\xi_{\textrm{P}}$ has zero mean and inherits the variance of the centroid estimator, which is given for a generalized radiant exitance profile $m(x\vert\phi,\theta)$ and point spread function $\psi(x)$ by (Appendix \ref{apx:estimators})
	\begin{equation}
	\sigma_{\textrm{P}}^2=\expval*{\Delta\hat{\phi}^2}=\frac{1}{n_1}\int\limits_{-\infty}^{\infty}x^2\Big(\abs{\psi(x)}^2+m(x\vert0,\theta)\Big)dx.
	\label{eq:centroid_variance}
	\end{equation}
	With i.i.d.~photon arrivals, the posterior distribution $p(\xi_{\textrm{P}})$ can be accurately approximated by a zero mean Gaussian distribution with variance $\sigma_{\textrm{P}}^2$ via the Central Limit Theorem. Since $\xi_{\textrm{P}}$ and $\xi_{\textrm{S}}$ are statistically independent random variables, the distribution $p(\xi)$ is given by 
	\begin{equation}
	p(\xi)=\bigg(\frac{1}{2\pi\big(\sigma_{\textrm{P}}^2+\sigma_{\textrm{S}}^2\big)}\bigg)^{1/2}\exp\bigg(-\frac{\xi^2}{2\big(\sigma_{\textrm{P}}^2+\sigma_{\textrm{S}}^2\big)}\bigg).
	\label{eq:p_xi}
	\end{equation}
	With a prior distribution on $\xi$ in hand, it is convenient to eliminate the non-random nuisance parameter $\phi$ from Eqs. \ref{eq:image_plane} and \ref{eq:direct_joint_probability} and reexpress them in terms of the random nuisance parameter $\xi$. Applying the change of variables $x_{i}'= x_{i}-\phi_{\textrm{0}}$ leads to
	\begin{align}
		&\Psi\big(x'_i\vert\xi,\theta\big)=\int_{-\infty}^{\infty}m(x\vert\xi,\theta)\abs{\psi(x_i-x)}^2dx.
		\label{eq:Stage1_probability}
		\\
		&P_{\textrm{D}}\big(\{x'_i\}_{i=1}^{n_1}\vert\xi,\theta\big)=\prod\limits_{i=1}^{n_1} \Psi\big(x'_i\vert\xi,\theta\big),
		\label{eq:Stage1_joint_probability}
	\end{align}
	where we have used the fact that $m(x\vert\phi,\theta)=m(x-\phi\vert,0,\theta)$ for an object with centroid $\phi$.  
	
	In the second stage of our receiver design, the remaining integration time is allocated to a BSPADE measurement which detects a Poisson distributed number of photons $n_2$ with mean $N_2=(1-\alpha)N$. After the changes of variables $x'=x-\phi_{\textrm{0}}$ and $s'=s-\phi_{\textrm{0}}$ in Eq. \ref{eq:BSPADE_model}, the probability of single photon detection in the $q^{\textrm{th}}$ spatial mode becomes
	\begin{multline}
		g(q\vert\xi,\theta)= \\
		\int\limits_{-\infty}^{\infty}m(x'\vert\xi,\theta)\abs{\int\limits_{-\infty}^{\infty}\psi(s'-x')f_q(s')ds'}^2 dx'
		\label{eq:Stage2_probability}
	\end{multline}
	so that the random parameter $\xi$ again replaces the non-random parameter $\phi$ as the nuisance parameter conditioning the measurement outcome distribution. In particular, a 0-BSPADE measurement, which uses a PSF-matched target mode $f_0(x)=\psi(x)$, is known to saturate the quantum limit for estimating sub-Rayleigh separation between two incoherent emitters when the mode sorter is perfectly aligned to the centroid \cite{Tsang2016a,Kerviche2017,Rehacek2017b,Tsang2018a}. Using 0-BSPADE as the second stage measurement, the target mode probability simplifies to
	\begin{equation}
	g(0\vert\xi,\theta)=\int\limits_{-\infty}^{\infty}m(x'\vert\xi,\theta)\abs{\Gamma(x')}^2dx'
	\label{eq:Stage2_0BSPADE_probability}
	\end{equation}  
	where $\Gamma(x)=\int_{-\infty}^{\infty}\psi(s-x)\psi(s)ds'$ is the autocorrelation function of the PSF. The joint probability distribution for the second stage (Eq. \ref{eq:BSPADE_joint}) then becomes
	\begin{equation}
	P_{\textrm{B}}(k,n_2 \vert\xi,\theta)=g(0\vert\xi,\theta)^{k}\big(1-g(0\vert\xi,\theta)\big)^{n_2-k},
	\label{eq:Stage2_joint_probability}
	\end{equation}
	where $k$ and $n_2$ constitute a sufficient statistic for a measurement which counts $k$ photons in the PSF-matched mode. Since the centroid estimate error $\xi_{\textrm{P}}$ and therefore also the total misalignment $\xi$ depend on the direct detection data, Eq. $\ref{eq:Stage2_joint_probability}$ is implicitly a conditional probability given the outcome of the first stage. 
	
	Finally, a digital post-processing step is performed to estimate $\theta$ from the data collected by both stages. With no \textit{a priori} information available on the parameter of interest, we construct an appropriate likelihood function and use a maximum likelihood estimator. The total joint probability distribution for both stages is simply the product of the outcome distribution for the first stage (Eq. \ref{eq:Stage1_joint_probability}) and that of the second stage conditioned on the first (Eq. \ref{eq:Stage2_joint_probability}). Since $\xi$ is a nuisance parameter with a known prior distribution $p(\xi)$, it can be removed in the construction of the likelihood function via marginal integration \cite{Berger1999}, resulting in the likelihood function
	\begin{multline}   
		P_{\textrm{2-Stage}}\big(\{x'_i\}^{n_1},k,n_2\vert\theta\big)= \\
		\int\limits_{-\infty}^{\infty}
		P_{\textrm{D}}\big(\{x'_i\}_{i=1}^{n_1}\vert\xi,\theta \big)P_{\textrm{B}}\big(k,n_2\vert\xi,\theta\big)
		p(\xi)d\xi.
		\label{eq:marginal_likelihood}
	\end{multline}
	We stress that since $\xi$ models static misalignment which is sampled once for the entire imaging trial, the marginal integral must be taken over the entire joint distribution and not over Eqs.  \ref{eq:Stage1_probability} or \ref{eq:Stage2_0BSPADE_probability}, because doing so would ignore the correlation present in the misalignment affecting each photon. Eq. \ref{eq:marginal_likelihood} takes into account the entire two-stage data set and can be used as a likelihood function to calculate the ML estimator
	\begin{equation}
	\hat{\theta}_{\textrm{2-Stage}}=\argmax_{\theta} P_{\textrm{2-Stage}}\big(\{x'_i\}^{n_1},k,n_2\vert\theta\big).
	\label{eq:ML_theta}
	\end{equation}
	
	We assume in our results a Gaussian apodized aperture along the image plane coordinate $x$, and we expect that our receiver design and results would not change dramatically if the more experimentally relevant circular aperture model were employed \cite{Kerviche2017}. The 1D coherent point spread function (PSF) for a Gaussian aperture is given by
	\begin{equation}
	\psi(x)=\bigg(\frac{1}{2\pi\sigma^2}\bigg)^{1/4}\textrm{exp}\bigg(\frac{-x^2}{4\sigma^2}\bigg),
	\label{eq:PSF}
	\end{equation}
	where the spatial parameter $\sigma$ is a characteristic PSF width which depends on the source center wavelength $\lambda$ and the numerical aperture (NA) of the imaging system as $\sigma=\lambda/(2\pi\textrm{NA})$. For distant objects, such as in astronomical imaging, $x$, $\phi$ and $\theta$ can be treated as observational angles and the PSF width is given by the angle $\sigma\approx\lambda/D$, where $D$ is the diameter of the entrance aperture of the imaging system (Fig. \ref{fig:2Stage}A). Here, we term any distance smaller than the PSF width $\sigma$ to be ``sub-Rayleigh." 
	
	For the specific case of estimating the separation between two point sources (Eq. \ref{eq:PointSources_radiance}) using a Gaussian aperture (Eq. \ref{eq:PSF}), the variance in the centroid estimate error becomes (Appendix \ref{apx:estimators})
	\begin{equation}
	\sigma_{\textrm{P}}^2=\expval*{\Delta\hat{\phi}^2}=\frac{\sigma^2}{n_1}\bigg(1+\frac{\theta^2}{4\sigma^2}\bigg),
	\end{equation}
	which depends on $\theta$ and $n_1$ and converges as $\theta\ll\sigma$ to the well known variance for localization of a single point source of $\sigma^2/n_1$. The direct detection single-photon outcome probability density is given by
	\begin{multline}   
		\Psi(x_i\vert\xi,\theta)=\frac{1}{2\sqrt{2\pi\sigma^2}} \\
		\Bigg[\exp\bigg(\frac{x_i-\xi+\theta/2}{2\sigma^2}\bigg)+\exp\bigg(\frac{x_i-\xi-\theta/2}{2\sigma^2}\bigg)\Bigg],
	\end{multline}
	while for the 0-BSPADE measurement matched to the Gaussian aperture, $\Gamma(x)=\exp(-x^2/4\sigma^2)$, and Eq. \ref{eq:Stage2_probability} simplifies to \cite{Tsang2016a}
	\begin{equation}
	g(0\vert\xi,\theta)=\frac{1}{2}\Big[\exp\big(-Q(\xi,\theta)\big)+\exp\big(-Q(\xi,-\theta)\big)\Big],
	\label{eq:Stage2_0BSPADE_probability_Gaussian}
	\end{equation}
	where $Q(\xi,\theta)=(\xi-\theta/2)^2/4\sigma^2$. Once the data from both stages is acquired, Eq. \ref{eq:ML_theta} can be evaluated numerically to estimate $\theta$. 
	
	\subsection{Adaptive Optimization of Two-Stage Allocation Ratio}
	\label{sec:alpha}
	
	Our receiver design requires a choice of the two-stage allocation ratio $\alpha$, a free design parameter which can be tuned for a given imaging trial to optimize estimation precision. Since BSPADE is a more information rich measurement than direct detection for sub-Rayleigh separation estimation \cite{Tsang2016a}, designating more time to the second stage can suppress the estimator variance. However, Eq. \ref{eq:centroid_variance} implies that reducing the signal integrated by direct detection comes at the cost of worsening alignment uncertainty and thereby weakening the effectiveness of the BSPADE measurement. For an experiment with finite mean photon number $N$, the optimal $\alpha$ will depend non-trivially on both the imaging system and object parameters.
	
	To optimize $\alpha$ for a given experiment, we seek to minimize the ML estimator variance. We write this variance as $\expval*{\Delta\hat{\theta}_{\textrm{2-Stage}}^2}=\big(\expval*{\Delta\hat{\theta}_{\textrm{D}}^2}^{-1}+\expval*{\Delta\hat{\theta}_{\textrm{B}}^2}^{-1}\big)^{-1}$, a harmonic sum of variances from the two constituent measurements, where $\hat{\theta}_{\textrm{D}}$ and $\hat{\theta}_{\textrm{B}}$ are ML estimators for a direct detection measurement and a 0-BSPADE measurement integrating $N_1$ or $N_2$ photons, respectively. The direct detection estimator variance can be reasonably approximated using the Cram\'er-Rao bound (CRB) $\mathcal{I}_{\textrm{D}}^{-1}(\theta)$, which is given in Appendix \ref{apx:estimators}. When the ML estimator is unbiased, the CRB provides a tight lower bound on its variance that scales inversely with the number of collected photons, such that $\expval*{\Delta\hat{\theta}_{\textrm{D}}^2}\approx\mathcal{I}_{\textrm{D}}^{-1}(\theta)/N_1$ \cite{VanTrees2013}. On the other hand, Cram\'er-Rao bounds are not suitable as approximations for the BSPADE estimator variance $\expval*{\Delta\hat{\theta}_{\textrm{B}}^2}$. Several variants of the Cram\'er-Rao bound have been developed for estimation tasks in the presence of random nuisance parameters which are sampled once for many i.i.d. measurements \cite{VanTrees2013,Bobrovsky1987,Miller1978,DAndrea1994,Noam2009}, but these bounds are too loose to be useful for minimization of the two-stage estimator variance. Instead, we resort to direct numerical evaluation of the BSPADE estimator variance $\expval*{\Delta\hat{\theta}_{\textrm{B}}^2}$, which is computationally tractable due to the simple structure of the BSPADE measurement (Appendix \ref{apx:estimators}). 
	
\begin{figure}[htb]
		\centering
		\includegraphics[width=\linewidth]{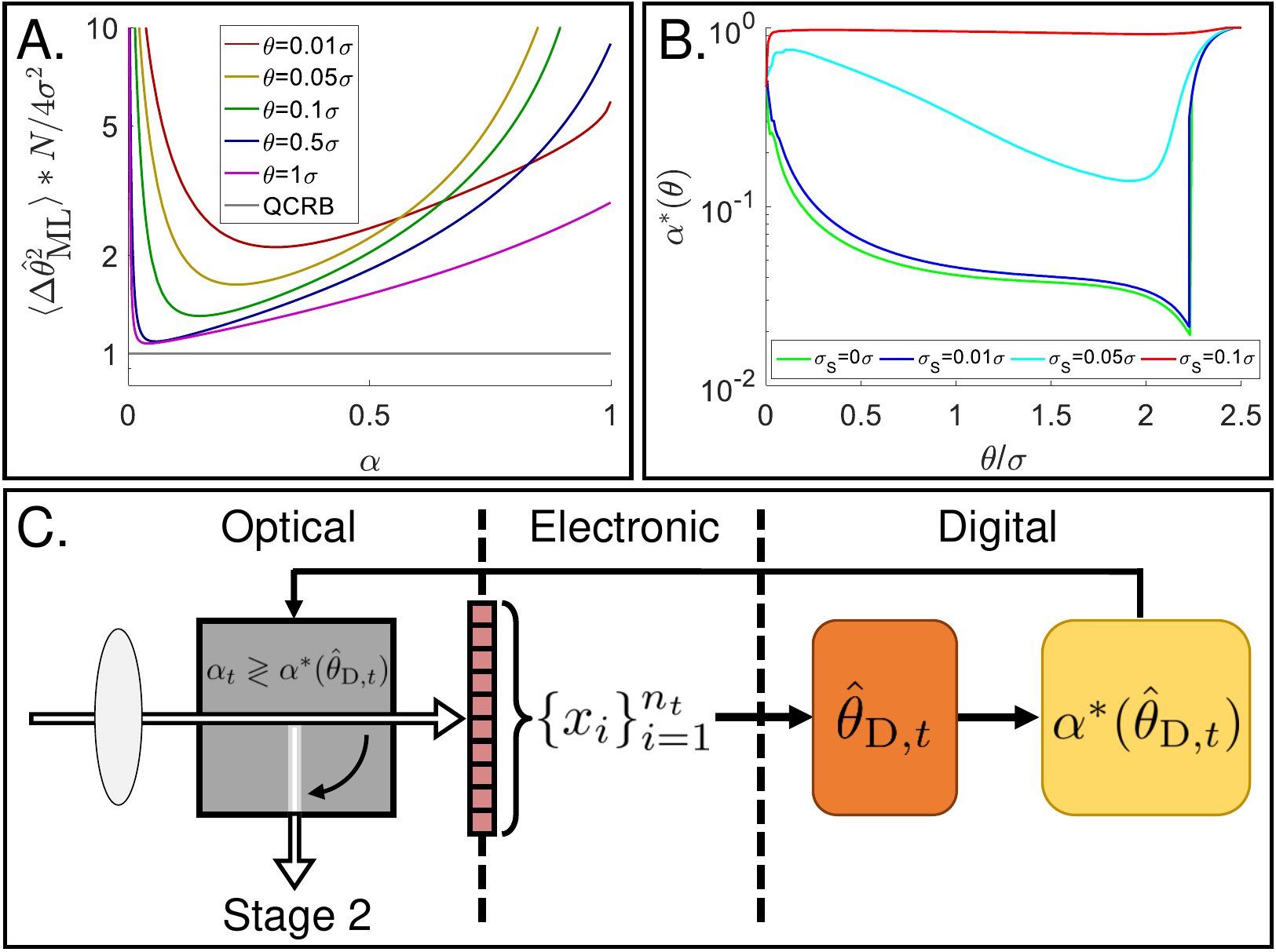}
		\caption{A. Variance of approximated ML estimator for two stage receiver as a function of the allocation parameter $\alpha$ when $N=100,000$. Note that the y axis is normalized to $N/4\sigma^2$, the ultimate lower bound for estimating the separation of two incoherent point sources \cite{Tsang2016a}. B. Optimal allocation ratio which minimizes the estimator variance $\expval*{\Delta\hat{\theta}^2}$ as a function of point source separation for different amounts of systematic misalignment ($N=100,000$). C. Adaptive method for optimizing the design parameter $\alpha$, with the an iteration of the optimization loop shown at time $t$. Separation estimation $\hat{\theta}_{\textrm{D},t}$ is performed using all photon arrival data recorded through time $t$.}
		\label{fig:opt_alpha}
	\end{figure}

	The optimal allocation ratio $\alpha^*(\theta)$ can then be found by minimizing $\expval*{\Delta\hat{\theta}_{\textrm{2-Stage}}^2}$ (Fig. \ref{fig:opt_alpha}A). Unfortunately, for the point source separation estimation task, this optimal value is a function of the unknown parameter of interest due to the $\theta$ dependence of both $\expval*{\Delta\hat{\theta}_{\textrm{D}}^2}$ and $\expval*{\Delta\hat{\theta}_{\textrm{B}}^2}$, as shown in Fig. \ref{fig:opt_alpha}B. While $\alpha^{\ast}(\theta)$ tends to $1/2$ in the limit of small point source separation, in agreement with recent results \cite{Zhou2019}, it varies significantly throughout the regime where $\theta<=2\sigma$. Furthermore, when $\theta$ grows beyond $2\sigma$, $\alpha^{\ast}(\theta)$ sharply increases to 1, suggesting that allocating the entire integration time to direct detection becomes optimal. This behavior reflects the fact that the estimation performance of 0-BSPADE degrades at large emitter separation due to decreasing overlap between the image-plane optical field and the PSF-matched target mode, such that eventually direct detection becomes the superior measurement for estimating the emitter separation \cite{Tsang2016a}. As a result, if no prior knowledge of $\theta$ is available, the chosen measurement allocation ratio may differ greatly from the optimal value. 
	
	Our two-stage receiver design is inherently equipped to overcome this problem via a modification which uses the data from the first stage to dynamically build up the requisite knowledge of $\theta$ needed to optimize $\alpha$. A summary of this adaptive measurement approach is shown in Fig. \ref{fig:opt_alpha}C. During the first measurement stage, a closed feedback loop periodically checks whether it is advantageous to continue with direct detection or to switch to the 0-BSPADE measurement. This decision is made by comparing $\alpha_t=t/T$, where $t$ is the current elapsed time of the measurement, against $\alpha^{\ast}(\hat{\theta}_{\textrm{D},t})$, where the emitter separation estimate $\hat{\theta}_{\textrm{D},t}$ is an approximate ML estimator (see Appendix \ref{apx:estimators}) for $\theta$ based on the $n_t$ photons detected by the FPA during the time $t$. While direct detection is a poor measurement for sub-Rayleigh emitter separation \cite{Tsang2016a}, the coarse information it provides can be used to calculate the optimal allocation parameter $\alpha^{\ast}(\hat{\theta}_{\textrm{D},t})$ in the absence of prior information on $\theta$. As soon as a loop instance returns $\alpha_t>\alpha^{\ast}(\hat{\theta}_{\textrm{D},t})$, the receiver applies an optical switch to route the collected light to the BSPADE measurement, and the allocation ratio for the experiment becomes $\alpha=\alpha_t$. If this switch is never signaled over the entire integration time, $\alpha=1$ and the receiver has simply implemented a direct detection measurement. 
	
	\subsection{Receiver Performance: Point Source Separation Estimation}
	\label{sec:estimation}
	We implemented our sequential measurement scheme in a Monte Carlo simulation to quantify its improvement over direct detection for the point source separation estimation task with an unknown centroid as a nuisance parameter. In each simulation of the two-stage receiver, $\alpha$ was either pre-determined to $.5$ or adaptively optimized scheme described above. We compare these results to direct detection, which is implemented using the exact same simulation code with $\alpha=1$. In all cases, the simulated receiver was given knowledge of the systematic misalignment parameter $\sigma_{\textrm{S}}$ but was given no prior information regarding the object parameters $\theta$ and $\phi$. 
	
	\begin{figure}[htbp]
		\centering
		\includegraphics[width=\linewidth]{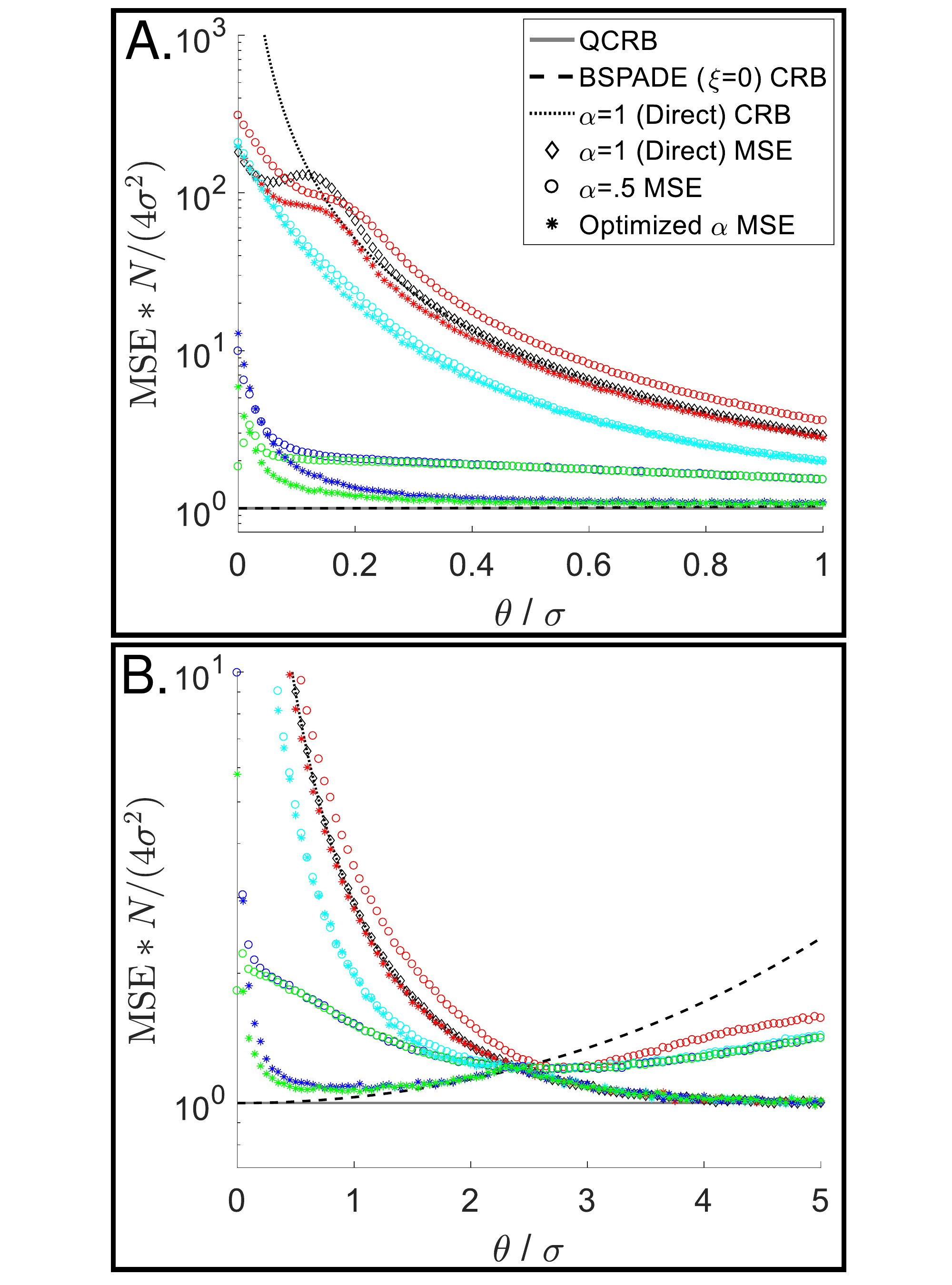}
		\caption{A. Monte Carlo simulation results for estimation of sub-Rayleigh emitter separation $\theta$ with total mean photon number $N=100,000$ compared with Cram\'er-Rao bounds. Black diamonds and dotted black line: simulation MSE and CRB for direct detection (i.e., two-stage receiver with $\alpha=1$). Circles: simulation MSE using $\alpha=0.5$. Stars: simulation MSE using adaptive optimization of $\alpha$. Colors signify systematic misalignment RMSE (red: $\sigma_s=0.1\sigma$, cyan: $\sigma_s=0.05\sigma$, blue: $\sigma_s=0.01\sigma$, green: $\sigma_s=0$). B. Same results extended to the super-Rayleigh regime ($\theta>\sigma$).}
		\label{fig:MSE_point_sources}
	\end{figure}
	
	Fig. \ref{fig:MSE_point_sources}A plots the simulated mean squared error (MSE) for sub-Rayleigh separation estimation as a function of the true value of $\theta$ for different amounts of random systematic misalignment.  First consider a \emph{standard} two-stage receiver, for which $\alpha$ is fixed at .5, in comparison with image-plane direct detection and with the quantum Cram\'er-Rao bound (QCRB), a lower bound on estimator variance that applies regardless of the chosen measurement scheme \cite{Helstrom1976}. With no systematic misalignment $\xi_{\textrm{S}}$, the MSE attained by the standard two-stage receiver beats that of direct detection by up to two orders of magnitude and remains at approximately a constant factor of two above the QCRB for most of the sub-Rayleigh regime. This latter observation indicates that the preliminary measurement succeeds in sufficiently driving down the misalignment contribution $\xi_{\textrm{P}}$ such that the BSPADE measurement, which detects half of the available photon budget, achieves near quantum-limited estimation precision. This performance is robust to a complete lack of prior knowledge of the object centroid and to small amounts of systematic misalignment which are within achievable tolerances for modern space telescope pointing systems \cite{Mandic2018,Sacks2018} and xy sample positioning stages for microscopy \cite{Instrumente}, although the estimation performance degrades as the systematic misalignment increases.
	
	Dynamic optimization of the measurement allocation ratio improves the two-stage receiver in several ways. First, in the absence of systematic misalignment, the achievable MSE is reduced from the 50/50 implementation by about a factor of two for most of the sub-Rayleigh regime, nearly saturating the QCRB and beating the direct detection MSE by up to two orders of magnitude. Only deep into the sub-Rayleigh regime does the adaptive scheme fail to improve estimation precision over the standard two-stage receiver, both because 0.5 is already close to the optimal value for $\alpha$ \cite{Zhou2019} and because the variance of the $\hat{\theta}_{\textrm{D},t}$ estimator becomes too large for direct detection to provide an adequate $\theta$ estimate for the optimization scheme. Note that even as $\theta$ approaches zero, the adaptive two-stage receiver outperforms direct detection by over an order of magnitude in MSE. Second, in the presence of systematic misalignment, the adaptive method at worst matches the estimation precision achieved with 50/50 allocation and even hedges against large systematic misalignment by capping the estimation MSE to that of direct detection. 
	
	Finally, adaptive optimization of $\alpha$ adds value in the important case where the emitter separation could range from sub-Rayleigh to super-Rayleigh. Fig. \ref{fig:MSE_point_sources}B shows that at larger emitter separation, where direct detection is a preferable measurement to even a perfectly aligned BSPADE \cite{Tsang2016a}, the adaptive two-stage receiver automatically reverts to allocating the entire integration time to direct detection. The MSE for the standard two-stage receiver, on the other hand, increases as the separation grows due to the deteriorating efficacy of the 0-BPSADE measurement. Our adaptive two-stage receiver enables near quantum optimal performance for this estimation task even when the receiver is given no prior constraints on the parameter of interest or the nuisance parameter. 
	
	\subsection{Receiver Performance: Extended Source Length Estimation}
	\label{sec:line_source}
	To demonstrate the versatility of our two stage receiver design, we additionally applied our receiver to estimating the length of a uniform extended object. The QCRB for this task was considered in recent work \cite{Dutton2019}, and a 0-BSPADE was found to be a quantum optimal measurement when the mode sorter is perfectly aligned to the centroid. In fact, both the separation between two point emitters and the length of a uniform extended object correspond to the second moment of the respective object distributions, for which BSPADE is optimal in the limit of small object length with perfect alignment \cite{Tsang2019} but negatively affected by misalignment \cite{Zhou2019}. 
	
	Assume an incoherent extended object with length $\theta$ and a uniformly distributed radiant exitance profile 
	\begin{equation}
	m(x\vert\phi,\theta)=\frac{1}{\theta}\textrm{Rect}\bigg[\frac{x-\phi}{\theta}\bigg],
	\label{eq:radiant_exitance_line}
	\end{equation}
	where the object centroid $\phi$ is again cast as a nuisance parameter for the estimation task. Assuming a Gaussian PSF, the data from the first measurement stage will be distributed with a spatial probability density of \cite{Dutton2019}
	\begin{align}
		&\Psi(x_i\vert\phi,\theta)=\int_{-1/2}^{1/2}\abs{\psi(x-\phi-\theta s)}^2ds \\
		&=\frac{1}{2\theta}\bigg[\erf\Big(\frac{x-\phi+\theta/2}{\sqrt{2}\sigma}\Big)-\erf\Big(\frac{x-\phi-\theta/2}{\sqrt{2}\sigma}\Big)\bigg].
		\label{eq:direct_line}
	\end{align}
	Using the approximate ML estimator $\hat{\phi}$ of the object centroid, the centroid error variance is  (Appendix \ref{apx:estimators})
	\begin{equation}
	\sigma_{\textrm{P}}^2=\expval*{\Delta\hat{\phi}^2}=\frac{\sigma^2}{n_1}\bigg(1+\frac{\theta^2}{12\sigma^2}\bigg).
	\label{eq:line_centroid_variance}
	\end{equation} 
	For a misaligned 0-BSPADE measurement matched to the Gaussian PSF, the single photon target mode probability is given by 
	\begin{multline}
		g(0\vert\xi,\theta)= \\
		\frac{\sqrt{\pi}\sigma}{\theta}\Big[\erf\big(\sqrt{Q(\xi,\theta)}\big)-\erf\big(\sqrt{Q(\xi,-\theta)}\big)\Big].
		\label{eq:line_Stage2_0BSPADE_probability}
	\end{multline}
	The maximum likelihood estimator $\hat{\theta}_{\textrm{ML}}$ can then be calculated using Eq. \ref{eq:ML_theta} in order to adaptively optimize $\alpha$. 
	
	\begin{figure}[htb]
		\centering
		\includegraphics[width=\linewidth]{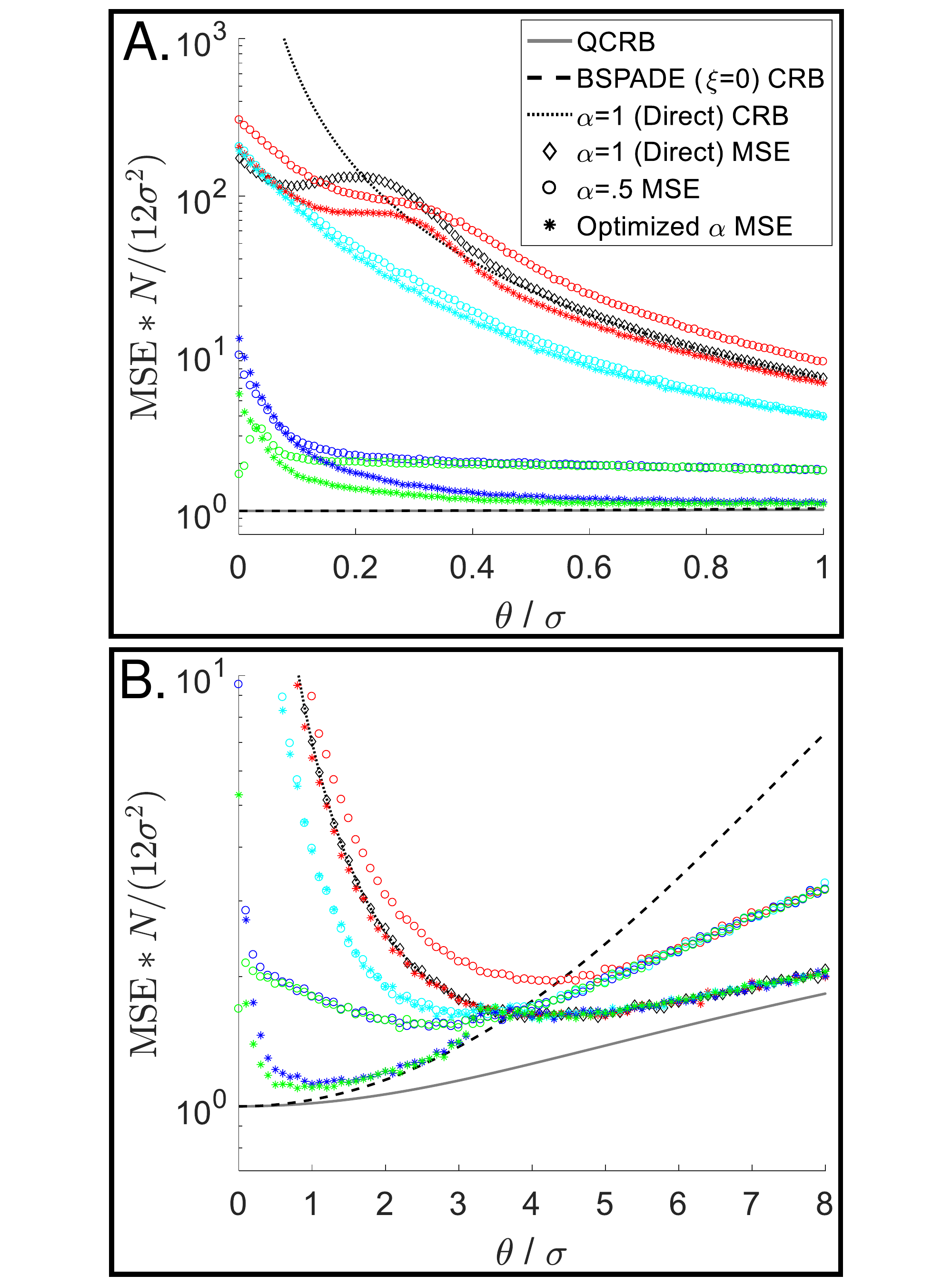}
		\caption{Monte Carlo simulation results for estimation of object length $\theta$ with total mean photon number $N=100,000$ compared with Cram\'er-Rao bounds. All plot features are identical to Fig. \ref{fig:MSE_point_sources} except for the vertical axis normalizations.}
		\label{fig:MSE_line_source}
	\end{figure}
	
	The results of Monte Carlo simulations for estimating the length of an extended source using the two-stage receiver are shown in Fig. \ref{fig:MSE_line_source}, and the two-stage receiver outperforms direct detection by a similar factor in the sub-Rayleigh regime as compared with emitter separation estimation. In addition, adaptive optimization of the free parameter $\alpha$ again improves the error by up to a factor of two for sub-Rayleigh imaging. The most significant difference from point source separation estimation is the behavior of the MSE results as the object length grows beyond the PSF width $\sigma$, when the MSEs for the two stage receiver increase at a faster rate than in the previous case. This is due to an overall decrease in length estimation precision as $\theta$ grows, typified by the increasing QCRB and explained by the fact that regions at the interior of the extended source provide less information about the object length compared with the edges \cite{Dutton2019}. Even so, the adaptive two-stage receiver remains near the quantum limit for length estimation for the entire range of possible values for $\theta$ with no prior knowledge of the object length or centroid.
	
	\section{Discussion}
	\label{sec:discussion}
	
	As the primary result of our work, we conclude that a two stage receiver design can overcome the problems associated with the sensitivity of spatial-mode-sorting based imaging receivers to optical misalignment. Our proposed solution implements a preliminary direct detection measurement which serves three purposes for the imaging receiver. First and most importantly, the preliminary measurement provides an estimate of the object centroid which allows for suppression of the optical misalignment associated with the BSPADE measurement. Second, the preliminary measurement provides crucial information which enables a real time decision of when to optimally switch to the second stage in the adaptive implementation of the receiver. Third, the direct detection data should be utilized in estimating the parameter of interest, since it becomes the predominant source of information on the parameter of interest in the super-Rayleigh regime for both tasks considered here.
	
	As an equally significant contribution from this work, we have introduced a framework for using adaptive sequential measurements which, by building up previously unavailable prior information about a scene, can enable operation of mode-sorting imaging receivers near the limits set by quantum mechanics. This framework could easily be expanded to the presence of additional nuisance parameters and multiple parameters of interest. For example, optical defocus along the axial dimension \cite{Prasad2019,Yu2018a}, three-dimensional rotation of the object \cite{Napoli2019}, and spatially non-uniform object brightness \cite{Rehacek2017a,Rehacek2018} are a few examples of additional complexities which have been shown to influence the quantum limits and optimal measurements for estimating the displacement between point sources. In a generalized scheme, preliminary measurements could adaptively pre-estimate each nuisance parameter in order to optimally prepare one or more spatial-mode-sorting measurements, potentially saturating estimation precision limits on the parameter(s) of interest. In some cases, direct detection would not be the preferred preliminary measurement; in fact the optimal measurement for estimating the centroid for the two estimation tasks described here is not direct detection \cite{Tsang2016a} and remains unknown. Furthermore, some quantitative imaging tasks may involve multiple viable hypotheses, one of which must to be chosen either prior to or alongside parameter estimation \cite{Lu2018}. Finally, future work on more complex imaging tasks could explore the value of repeatedly switching back and forth between measurement stages, a feature which appears in quantum sensing techniques \cite{Dolinar1973, Zhuang2017}. We emphasize that our sequential estimation procedure could have easily been formulated in the Bayesian framework, and moving to a Bayesian estimation scheme will likely be fruitful for incorporating multiple adaptive measurements as well as partial prior information on object parameters. 
	
	\begin{acknowledgments}
		The authors would like to thank Mankei Tsang, Jelena Vuckovic, Rahul Trivedi, and Daniil Lukin for insightful discussions. This work was supported by a DARPA EXTREME program seedling under Contract No. HR0011-17-1-0007. This document does not contain technology or technical data controlled under either the U.S. International Traffic in Arms Regulations or the U.S. Export Administration Regulations.
	\end{acknowledgments}
	
	\appendix
	\section{Estimators for the Two-Stage Imaging Receiver}
	\label{apx:estimators}
	
	The first estimator that is needed for our two stage receiver design is an estimator $\hat{\phi}$ for an object centroid from ideal direct detection data which is used to align the BSPADE receiver. To implement a maximum likelihood estimator for a generalized 1D incoherent object with image-plane intensity distribution $\Psi(x_i\vert\phi,\theta)$ with an unknown transverse parameter $\theta$, we use the joint photon arrival probability distribution $P_{\textrm{D}}\big(\{x_i\}_{i=1}^{n_1}\vert\phi,\theta\big)=\prod_{i=1}^{n_1}\Psi(x_i\vert\phi,\theta)$ as a likelihood function. The photon arrival positions $x_i$ can registered to the object centroid, such that the arbitrary single photon arrival probability distribution can be written $\Psi(x_i-\phi\vert0,\theta)=\Psi(x_i\vert\phi,\theta)$. Taking the derivative of the log-likelihood function with respect to $\phi$ gives 
	\begin{multline}
		\frac{\partial}{\partial\phi} \log\Big[P_{\textrm{D}}\big(\{x_i\}_{i=1}^{n_1}\vert\phi,\theta\big)\Big]= \\
		\sum\limits_{i=1}^{n_1}\frac{\partial}{\partial\phi}\log\big[\Psi(x_i-\phi\vert0,\theta)\big].
	\end{multline}
	Setting this expression equal to zero to find an extremum achieves generalized ML estimation for the centroid. However, this approach is undesirable for centroid estimation because it cannot be expressed as an explicit estimator and because it depends on the unknown parameter $\theta$. We can find an approximate form for an explicit estimator by placing a further constraint on the object such that $\lim_{\theta\to 0} \Psi(x\vert0,\theta)=\abs{\psi(x)}^2$, restricting the object distribution to even functions which limit to a point source as the object scale is reduced to zero. Applying this limit gives 
	\begin{equation}
	0=\sum\limits_{i=1}^{n_1} \frac{\frac{\partial}{\partial\phi}\abs{\psi(x_i-\phi)}^2}{\abs{\psi(x_i-\phi)}^2},
	\label{eq:implicit_centroid_estimator}
	\end{equation}
	which for a Gaussian PSF $\psi(x)$ (Eq. \ref{eq:PSF}) is equivalent to the explicit ``center of mass" estimator
	\begin{equation}
	\hat{\phi}=\frac{1}{n_1}\sum\limits_{i=1}^{n_1} x_i.
	\label{eq:COM_estimator}
	\end{equation}
	
	To find the estimator variance $\expval*{\Delta\hat{\phi}}$, we use the well known result that the variance of estimator given in Eq. \ref{eq:COM_estimator} scales inversely with the number of i.i.d. events, with a scaling factor given by the second central moment of the single event probability distribution. For the single photon arrival distribution $\Psi(x\vert\phi,\theta)$, we can set $\phi=0$ in the variance calculation because of the shift invariance property of ideal direct detection. Therefore,
	\begin{align}
		\expval*{\Delta\hat{\phi}}&=\frac{1}{n_1}\int\limits_{-\infty}^{\infty}x^2\Psi(x\vert0,\theta)dx 
		\label{eq:COM_variance1} \\
		&=\frac{1}{n_1}\int\limits_{-\infty}^{\infty}m(s\vert0,\theta)\int\limits_{-\infty}^{\infty}x^2\abs{\psi(x-s)}^2dx ds 
		\label{eq:COM_variance2} \\
		&=\frac{1}{n_1}\int\limits_{-\infty}^{\infty}m(s\vert0,\theta)\int\limits_{-\infty}^{\infty}(u+s)^2\abs{\psi(u)}^2du ds
		\label{eq:COM_variance3} \\
		&=\frac{1}{n_1}\Bigg[\int\limits_{-\infty}^{\infty}u^2\abs{\psi(u)}^2du+\int\limits_{-\infty}^{\infty}s^2m(s\vert0,\theta)ds\Bigg],
		\label{eq:COM_variance5}
	\end{align} 
	where the substitution $u=x-s$ is made in Eq. \ref{eq:COM_variance3}, and the cross term involving the factor $2us$ from Eq. \ref{eq:COM_variance3} is ignored because the first moments of both $m(s\vert0,\theta)$ and $\abs{\psi(u)}^2$ are zero. The first term in Eq. \ref{eq:COM_variance5} depends only on the PSF and becomes $\int_{-\infty}^{\infty}u^2\abs{\psi(u)}^2du=\sigma^2$ for a Gaussian PSF (Eq. \ref{eq:PSF}). The second term, which depends only on the object radiant exitance distribution, is easily found to be $\int_{-\infty}^{\infty}s^2m(s\vert0,\theta)ds=\theta^2/4$ for the two point source object and $\int_{-\infty}^{\infty}s^2m(s\vert0,\theta)ds=\theta^2/12$ for the extended object, leading to Eqs. \ref{eq:centroid_variance} and \ref{eq:line_centroid_variance}, respectively. 
	
	The ultimate goal of the two-stage receiver design is to estimate the parameter of interest $\theta$, which can be achieved by numerical evaluation of Eq. \ref{eq:ML_theta}. For the adaptive version of the two stage receiver, however, a further requirement is a calculation of the two-stage receiver estimator variance $\expval*{\Delta\hat{\theta}_{\textrm{2-Stage}}^2}=\big(\expval*{\Delta\hat{\theta}_{\textrm{D}}^2}^{-1}+\expval*{\Delta\hat{\theta}_{\textrm{B}}^2}^{-1}\big)^{-1}$ to optimize the measurement allocation parameter $\alpha$. The direct detection estimator variance can be approximated by the Cram\'er-Rao bound, which is calculated as the inverse of the $\theta$ dependent Fisher information, given by \cite{VanTrees2013}
	\begin{equation}
	\mathcal{I}_{\textrm{D}}(\theta)=\int_{-\infty}^{\infty}\frac{\big(\partial\Psi(x\vert 0,\theta)/\partial\theta\big)^2}{\Psi(x\vert 0,\theta)}dx,
	\label{eq:CFI}
	\end{equation} 
	where the centroid can be set to zero in this expression because of the shift invariant property of direct detection. For the two point source object with Gaussian PSF \cite{Yang2017a},
	\begin{multline}
		\mathcal{I}_{\textrm{D}}(\theta)= \frac{1}{4\sigma^2}-\\
		\frac{1}{2\sqrt{2\pi\sigma^5}}\int_{\infty}^{\infty}\frac{x^2\exp\Big(-\frac{(x+\theta/2)^2}{2\sigma^2}\Big)}{1+\exp\Big(-\frac{x\theta}{\sigma^2}\Big)}dx
	\end{multline}
	while for the extended source,
	\begin{multline}
		\mathcal{I}_{\textrm{D}}(\theta)=-\frac{1}{\theta^2}+\\
		\int_{-\infty}^{\infty}\frac{\Big[\exp\Big(\frac{-(x+\theta/2)^2}{2\sigma^2}\Big)+\exp\Big(\frac{-(x-\theta/2)^2}{2\sigma^2}\Big)\Big]^2}{4\pi\sigma^2\theta\Big[\erf\Big(\frac{x+\theta/2}{\sqrt{2\sigma^2}}\Big)-\erf\Big(\frac{x-\theta/2}{\sqrt{2\sigma^2}}\Big)\Big]}dx
	\end{multline}
	
	Conveniently, ML estimation is simplified by the limitation of the BSPADE measurement to just two single photon outcomes. If the misalignment parameter $\xi$ is known, the ML estimator for a 0-BSPADE measurement which integrates $n_2$ photons is simply given by the solution to the expression $g(0\vert\xi,\theta)=k/n_2$, where $g(0\vert\xi,\theta)$ is the single photon target mode probability. The ML estimator can be numerically solved or approximated in the small $\theta$ regime by taking a Taylor expansion of $g(0\vert\xi,\theta)-k/n_2$. For the results reported in the main text, we implement an approximation to third order in $Q(0,\theta)=\theta^2/16\sigma^2$ about $Q(0,\theta)=0$,
	\begin{multline}
		g(0\vert\xi,\theta)=c_0(\xi)+c_1(\xi)Q(0,\theta)+ \\
		c_2(\xi)Q(0,\theta)^2+c_3(\xi)Q(0,\theta)^3+O(\theta^8),
		\label{eq:ML_expansion}
	\end{multline}
	where $c_j$ is the $j^{\textrm{th}}$ expansion coefficient and there are no odd terms in $\theta$ because $\theta\geq0$ implies that the target mode probability is even with respect to $\theta$. Solving the cubic polynomial in Eq. \ref{eq:ML_expansion} results in the $\xi$ dependent estimator
	\begin{equation}
	\begin{aligned}
	\hat{\theta}_{\textrm{B}}(\xi)&=4\sigma\sqrt{\frac{1}{3}\Big(w^{1/3}-a_2(\xi)+\big(3a_1(\xi)-a_2(\xi)^2\big)w^{-1/3}\Big)} \\
	w&=\frac{1}{2}\Big(\sqrt{v^2+4\big(3a_1(\xi)-a_2(\xi)^2\big)^3}+v\Big) \\
	v&=-2a_2(\xi)^3+9a_1(\xi)a_2(\xi)-27a_0(\xi),
	\end{aligned}
	\end{equation}
	where $a_j(\xi)=c_j(\xi)/c_3(\xi)$, $j=0,1,2$. For the two emitter object we have
	\begin{equation}
	\begin{aligned}
	c_0(\xi)&=g(0\vert\xi,0)-\frac{k}{n_2} \\
	c_1(\xi)&=g(0\vert\xi,0)\big(2Q(\xi,0)-1\big) \\
	c_2(\xi)&=\frac{1}{6}g(0\vert\xi,0)\big(4Q(\xi,0)^2-12Q(\xi,0)+3\big) \\
	c_3(\xi)&=\frac{1}{90}g(0\vert\xi,0)\big(8Q(\xi,0)^3-60Q(\xi,0)^2 \\
	&+90Q(\xi,0)-15\big),
	\end{aligned}
	\end{equation}
	while for the extended object we have
	\begin{equation}
	\begin{aligned}
	c_0(\xi)&=g(0\vert\xi,0)-\frac{k}{n_2} \\
	c_1(\xi)&=\frac{1}{3}g(0\vert\xi,0)\big(2Q(\xi,0)-1\big) \\
	c_2(\xi)&=\frac{1}{30}g(0\vert\xi,0)\big(4Q(\xi,0)^2-12Q(\xi,0)+3\big) \\
	c_3(\xi)&=\frac{1}{630}g(0\vert\xi,0)\big(8Q(\xi,0)^3-30Q(\xi,0)^2 \\
	&+90Q(\xi,0)-15\big).
	\end{aligned}
	\end{equation}
	
	Since $\xi$ is a random parameter for the quantitative imaging tasks here, we use the $\xi$ independent estimator $\hat{\theta}_{\textrm{B}}=\int_{-\infty}^{\infty}\hat{\theta}_{\textrm{B}}(\xi)p(\xi)d\xi$, the weighted average of $\xi$ dependent estimators, as an approximation of the ML estimator for $\theta$. The variance of this approximate estimator as a function of $\theta$ is given by
	\begin{equation}
	\expval*{\Delta\hat{\theta}_{\textrm{B}}}=\sum\limits_{k=0}^{n_2}(\hat{\theta}_{\textrm{B}}-\theta)^2P_{\textrm{B}}(k,n_2 \vert\theta),
	\end{equation}
	where $P_{\textrm{B}}(k,n_2 \vert\theta)=\int_{-\infty}^{\infty}P_{\textrm{B}}(k,n_2 \vert\xi,\theta) p(\xi)d\xi$. Since there are exactly $n_2+1$ possible outcomes for a BSPADE measurement of $n_2$ photons, the joint probability for each value of $k$ can be exactly calculated with a finite number of evaluations of $\hat{\theta}_{\textrm{B}}$ and $P_{\textrm{B}}(k,n_2\vert\theta)$. 
	
	Finally, the adaptive two stage receiver requires a coarse estimate of the unknown parameter $\theta$ using a direct detection measurement which acts for time $t$ and detects $n_t$ photons. Taking the derivative of the direct detection log-likelihood function with respect to $\theta$ gives
	\begin{multline}
		\frac{\partial}{\partial\theta} \log\Big[P_{\textrm{D}}\big(\{x_i\}_{i=1}^{n_t}\vert\phi,\theta\big)\Big]= \\
		\sum\limits_{i=1}^{n_t}\frac{\partial}{\partial\theta}\log\big[\Psi(x_i-\phi\vert0,\theta)\big],
		\label{eq:ML_theta_implicit}
	\end{multline}
	where an implicit estimator can be found by setting Eq. \ref{eq:ML_theta_implicit} equal to zero. Since $\theta$ is constrained to be positive, $\Psi(x\vert0,\theta)$ must be even with respect to $\theta$ and  $\partial\log\big[\Psi(x\vert0,\theta)\big]/\partial\theta$ must be odd. Thus, taking a Taylor expansion about $\theta=0$ gives
	\begin{equation}
	\frac{\partial}{\partial\theta}\log\big[\Psi(x\vert0,\theta)\big]=c_1(x)\theta+c_3(x)\theta^3+O(\theta^5),
	\end{equation}
	where $c_1(x)$ and $c_3(x)$ are the first and third order Taylor expansion coefficients. An approximation for the explicit ML estimator is then
	\begin{equation}
	\hat{\theta}_{\textrm{D},t}\approx 
	\begin{cases}
	\sqrt{\frac{-\sum\limits_{i=1}^{n_t}c_1(x_i-\hat{\phi})}{\sum\limits_{i=1}^{n_t}c_3(x_i-\hat{\phi})}},
	& \textrm{if} \; \sum\limits_{i=1}^{n_t}
	c_1(x_i-\hat{\phi})<0
	\\
	0,              & \text{otherwise}
	\end{cases}
	\label{eq:Direct_ML_approx}
	\end{equation}
	which uses Eq. \ref{eq:COM_estimator} as a centroid estimate and where the first moment of $\partial\log\big[\Psi(x\vert0,\theta)\big]/\partial\theta$ is required to be negative to ensure the solution is a maximum. Using the appropriate Taylor expansions, the approximate separation estimator for the two emitter object becomes 
	\begin{equation}
	\hat{\theta}_{\textrm{D},t}\approx 
	2\sqrt{3}\sigma\sqrt{\frac{-\sum\limits_{i=1}^{n_t}\Big\{1-\big(\frac{x_i-\hat{\phi}_t}{\sigma}\big)^2\Big\}}{\sum\limits_{i=1}^{n_t}\big(\frac{x_i-\hat{\phi}_t}{\sigma}\big)^4}},
	\label{eq:points_Direct_ML_approx}
	\end{equation}
	while the approximate length estimator for the uniform extended object becomes
	\begin{equation}
	\hat{\theta}_{\textrm{D},t}\approx 
	2\sqrt{15}\sigma\sqrt{\frac{-\sum\limits_{i=1}^{n_t}\Big\{1-\big(\frac{x_i-\hat{\phi}_t}{\sigma}\big)^2\Big\}}{\sum\limits_{i=1}^{n_t}\Big\{\big(\frac{x_i-\hat{\phi}_t}{\sigma}\big)^4+4\big(\frac{x_i-\hat{\phi}_t}{\sigma}\big)^2-2\Big\}}},
	\label{eq:line_Direct_ML_approx}
	\end{equation}
	where for both cases a value of zero is assigned to the estimator if $(1/n_t)\sum_{i=1}^{n_t}\big((x_i-\hat{\phi}_t)/\sigma\big)^2<1$, following Eq. \ref{eq:Direct_ML_approx}.
	If some \textit{a priori} knowledge of $\theta$ is available, a Bayesian updating scheme could be used instead which takes into account both the prior information and the acquired data to perform the optimization.
	
	\bibliographystyle{unsrt}
	\bibliography{Adaptive_Two_Stage,Spatial_Mode_Sorting}
\end{document}